\documentclass[aps,showkeys,superscriptaddress,10.5pt,tightenlines]{revtex4}
\usepackage{times}
\usepackage{amsmath}
\usepackage{amsfonts}
\usepackage{graphicx}
\usepackage{epsfig}
\usepackage{dcolumn}
\usepackage{bm}
\usepackage{color}
\usepackage{longtable}
\usepackage{array}
\usepackage{multirow}

\newcommand{\pip}{\pi^+}
\newcommand{\pim}{\pi^-}
\newcommand{\piz}{\pi^0}

\newcommand{\psip}{\psi(2S)}

\newcommand{\jpsi}{J/\psi}

\newcommand{\EE}{e^+e^-}

\newcommand{\pp}{\pi^+\pi^-}

\newcommand{\lrt}{\leftrightarrow}
\newcommand{\rt}{\rightarrow}

\newcommand{\ppjpsi}{\pi^+\pi^- J/\psi}

\newcommand{\beq}{\begin{equation}}
\newcommand{\eeq}{\end{equation}}
\newcommand{\bitm}{\begin{itemize}}
\newcommand{\eitm}{\end{itemize}}
\newcommand{\LLB}{\Lambda\bar{\Lambda}}
\newcommand{\Lc}{\Lambda_{c}^+}
\newcommand{\Ds}{D_{s}^+}
\newcommand{\Dsstr}{D_{s}^{*+}}
\newcommand{\XmXmB}{\Xi^-\bar{\Xi^+}}
\newcommand{\XzXzB}{\Xi^0\bar{\Xi^0}}
\newcommand{\ALCP}{A_{CP}^{\Lambda}}
\newcommand{\aLam}{\alpha_{\Lambda}}
\newcommand{\aLamb}{\bar{\alpha}_{\Lambda}}
\newcommand{\AL}{\langle\alpha_{\Lambda}\rangle}
\newcommand{\AXimCP}{A_{CP}^{\Xi^-}}
\newcommand{\AXim}{\langle\alpha_{\Xi^-}\rangle}
\newcommand{\AXizCP}{A_{CP}^{\Xi^0}}
\newcommand{\AXiz}{\langle\alpha_{\Xi^0}\rangle}
\newcommand{\thetac}{\theta_C}
\newcommand{\sinthetac}{\sin\theta_C}

\newcommand{\ECM}{E_{\rm CM}}

\begin{document}
\title{New Physics Searches at the BESIII Experiment}
\author{Shenjian Chen}
 \email{sjchen@nju.edu.cn}
 \affiliation{School of Physics, Nanjing University,
 Nanjing 210093, China}
\affiliation{Nanjing Proton Source Research and Design Center,
 Nanjing 210093, China}
\author{Stephen Lars Olsen}
\email{solsensnu@gmail.com}
 \affiliation{University of Chinese Academy of Science,
 Beijing 100049, China}
 \affiliation{Institute for Basic Science,
 Daejeon 34126, South Korea}

\begin{abstract}

The Standard Model (SM) of particle physics, comprised of the unified electro-weak (EW) and Quantum
Chromodynamic (QCD) theories, accurately explains almost all experimental results related to the
micro-world, and has made a number of predictions for previously unseen particles, most notably the
Higgs scalar boson, that were subsequently discovered. As a result, the SM is currently universally
accepted as the theory of the fundamental particles and their interactions. However, in spite of its
numerous successes, the SM has a number of apparent shortcomings including: many free parameters that
must be supplied by experimental measurements; no mechanism to produce the dominance of matter over
antimatter in the universe; and no explanations for gravity, the dark matter in the universe, neutrino
masses, the number of particle generations, etc. Because of these shortcomings, there is considerable
incentive to search for evidence for new, non-SM physics phenomena that might provide important clues
about what a new, beyond the SM theory (BSM) might look like.  Although the center-of-mass energies
that BESIII can access are far below the energy frontier, searches for new, BSM physics are an important
component of its research program. Here we describe ways that BESIII looks for signs of BSM physics by
measuring rates for processes that the SM predicts to be forbidden or very rare, searching for non-SM
particles such as dark photons, making precision tests of SM predictions, and looking for violations
of the discrete symmetries $C$ and $CP$ in processes for which the SM-expectations are immeasurably small.  

\end{abstract}


\maketitle

\section{Introduction}
\label{Sec:intro}

The Standard Model consistently predicts the results of experimental measurements
and has emerged as the only viable candidate theory for describing elementary particle
interactions~\cite{Weinberg:2018apv}. In spite of its great success, there are a number
of reasons to believe that the Standard Model (SM) is not the ultimate theory, including:
\begin{itemize}
  \item The SM has 19 free parameters that must be supplied by experimental
    measurements. These include the quark, lepton and Higgs masses, the mixing angles of the
    Cabibbo-Kobayashi-Maskawa (CKM) quark-flavor mixing matrix, and the couplings of the electric,
    weak and QCD color forces.

  \item As first pointed out by Sakharov~\cite{Sakharov:1967dj}, the matter-antimatter asymmetry
    of the universe implies the existence of sizable $CP$-violating interactions in nature.
    However, The established SM mechanism for $CP$ violation fails to explain the matter-dominated universe
    by about ten orders of magnitude; there must be additional $CP$ violating mechanisms in nature
    beyond those contained in the SM.

  \item The model has no explanation for dark matter, which is, apparently, the dominant component
    of the mass of the universe.

  \item The particles in the SM are arranged in three generations of colored quarks and three
    generations of leptons; particle interactions are mediated by three forces, the color,
    electromagnetic and weak forces. The theory provides no explanation for why the number of
    generations is three and it does not account in any way for gravity, the fourth force that is
    known to exist.

\end{itemize}
As a result, there have been a huge number of experimental efforts aimed finding ``new physics,'' which
refers to new physical phenomena beyond the Standard Model (BSM) of particle physics. This may be, for
example, a new fundamental particle, such as a fourth generation quark or lepton, or
a new fundamental force carrier, such as a dark photon, high-mass gauge boson, a new Higgs-like
meson, etc.  Searches for new physics can be performed in two ways.  One method is to look
for direct production of new particles in collisions at high energy accelerators, for example at
the Large Hadron Collider, and reconstruct it from its SM decay products. Another way is to measure
precisely a decay process that can be accurately described by the SM, and look for deviations from
the SM prediction of the decay rate. According to quantum field theory (QFT), new heavy particles
can contribute to the decay process through virtual loop diagrams. These make precision measurements
sensitive to new physics, and this technique is widely used in high intensity collider experiments
such as BESIII~\cite{Prasad:2019ris,Wang:2015hdf,Godang:2013bb}.

Here we review highlights of some of these activities at BESIII.

\section{Rare Processes}

\subsection{\boldmath Search for flavor changing neutral currents {(FCNC)}}
\label{Sec:FCNC}

Flavor changing neutral current (FCNC) processes transform an up-type ($u, c, t$) or down-type ($d, s, b$)
quark into another quark of the same type but with a different flavor. In the SM, these processes are
mediated by the $Z$ boson and are known as neutral currents. However, they are strongly suppressed by
the Glashow–Iliopoulos–Maiani (GIM) cancellation~\cite{Glashow:1970gm} and only occur as second-order
loop processes. In many extensions of the SM, virtual TeV-scale particles can contribute competing
processes that lead to measurable deviations from SM-inferred transition rates or other properties.
Hence studies of rare FCNC processes are suitable probes for new physics. 

Recently, hints of discrepancies have been observed in the semi-leptonic FCNC processes of the $b$-quark,
$b\to s\ell^+\ell^-$~($\ell=e,\mu $) by the LHCb experiment~\cite{Capriotti:2018juw}:
(1) The differential branching fractions measured as a function of the squared four-momentum transferred
to the two leptons, $Q^2$, for several $B$-meson decay modes are below the theoretical
predictions~\cite{Aaij:2014pli,Aaij:2015esa,Aaij:2016flj,Aaij:2015xza,Detmold:2016pkz}. The largest
local discrepancy is a $3.3\sigma$ difference in the rate for $B_s^0\to\phi\mu^+\mu^-$ decay from its
SM-predicted value.
(2) The ratios of branching fractions for decays involving muons and electrons, defined as
$R_K= \frac{\mathcal{B}(B^+\to K^+\mu^+\mu^-)}{\mathcal{B}(B^+\to K^+e^+e^-)}$ and
$R_{K^*}= \frac{\mathcal{B}(B^+\to K^{*+}\mu^+\mu^-)}{\mathcal{B}(B^+\to K^{*+}e^+e^-)}$, which are unity
in the SM (i.e. lepton-flavor universality), were measured to be~\cite{Aaij:2014ora,Aaij:2017vbb}
\footnotesize
\begin{align*}
&R_K= 0.745^{+0.090}_{-0.074}\pm 0.036
	\text{ at central }Q^2\in[1.0, 6.0]\, \text{GeV}/c^2
	&& 2.6\sigma
	\\
&R_{K^*} = 0.66^{+0.11}_{-0.07}\pm 0.03
	\text{ at low }Q^2\in[0.045, 1.1]\, \text{GeV}/c^2
	&& 2.1\sigma-2.3\sigma
	\\
&R_{K^*} = 0.69^{+0.11}_{-0.07}\pm 0.05
	\text{ at central }Q^2\in[1.1, 6.0]\, \text{GeV}/c^2
	&& 2.4\sigma-2.5\sigma
\end{align*}
\normalsize
where the levels of deviations from the SM predictions are indicated.
(3) Measurements of the quantity $P'_5$, which is the chiral asymmetry produced by the interference
between the transversely and longitudinally polarized amplitudes in the decay $B\to K^{*+}\ell^+\ell^-$,
are $2.8\sigma$ and $3.0\sigma$ lower than the SM prediction in two $Q^2$ intervals below the $J/\psi$
resonance mass~\cite{Aaij:2015oid}.  Since these discrepancies could be evidence for new particles that
would extend the SM, it is important to check if there are similar deviations in the charm sector. 

While SM rates for FCNC transitions in the down-type $b$- or $s$-quark sectors are relatively frequent
because of the large mass of the top quark contribution to the loop, those in the up-type $c$-quark
sector are especially rare due to the small masses of the intermediate down-like quarks in the loop
that result in a strong GIM cancellation. For $c\to u$ transition rates for charmed and charmonia
particles that proceed via the SM loop contribution, dubbed as short distance (SD) effects, the expected
branching fractions are typically between
$< 10^{-8}$~\cite{Greub:1996wn,Fajfer:2001ad,Burdman:2001tf,Fajfer:2001sa,Paul:2011ar,Cappiello:2012vg}
and $10^{-10}$-$10^{-14}$~\cite{SanchisLozano:1993ki,Wang:2007ys}, respectively. For FCNC decays of charmed
mesons, the measured rates are enhanced by a few orders of magnitude by SM contributions from long
distance (LD) effects that proceed via di-lepton decays of ordinary $\rho$,~$\omega$~and~$\phi$~vector
mesons~\cite{Paul:2011ar,Cappiello:2012vg}. However, some extensions to the SM further enhance these FCNC
processes, sometimes by orders of
magnitude~\cite{Prelovsek:2000xy,Paul:2010pq,Fajfer:2001sa,Hill:1994hp,Aulakh:1982yn,Glashow:1976nt}.

The BESIII experiment has searched for $c$-quark FCNC processes in both charmed meson and charmonium
decays. No significant signals for new physics are found in any of the investigated decay modes, and
the inferred  90\% confidence level (CL) upper limits on the branching fractions
are summarized in Table~\ref{tab:FCNC}. 
\begin{itemize}
\item For the $D^0\to \gamma\gamma$ mode, the upper limit is consistent with that previously set by the
  BaBar experiment~\cite{Lees:2011qz}. The BESIII result is the first experimental study of this decay
  that uses $D^0$ mesons produced at the open-charm threshold. 
\item For the rare decays $D\to h(h^{(')}) e^+e^-$, where $h$=light meson(s), searches for four-body
  decays of $D^+$ mesons are performed for the first time, and the upper limits for $D^0$ meson decays
  are, in general, one order of magnitude better than  previous measurements~\cite{Patrignani:2016xqp}.
\item Searches for the FCNC decays $\psip\to D^0 e^+e^-$ and $\psip\to \Lambda_c^+\bar{p} e^+e^-$
  are performed for the first time. The upper limit on $J/\psi\to D^0 e^+e^-$ is two orders of magnitude
  more stringent than the best previous result, which was set by the BESII
  collaboration~\cite{Ablikim:2006qt}. 
\end{itemize}

\begin{table}[htbp]
\footnotesize
\caption{\footnotesize Results for the upper limit at 90\% CL on the branching fractions for various
  FCNC process searches performed at BESIII. Also listed are the best previous results.}
\begin{tabular}{lllll}
\hline\hline
Mode 	&Data	&$\mathcal{B}^\text{UL}$ at 90\% CL\; 	&Ref.$\;\;$ &Previous best $\mathcal{B}^\text{UL}$\\
\hline
$D^0\to\gamma\gamma$
	&$2.93\,\text{fb}^{-1}\,\psi(3770)\;\;$~~~~
	&$3.8\times 10^{-6}$	
	&\cite{Ablikim:2015djc}
	&$2.2\times 10^{-6}$
	\\
	\hline
$D^+\to\pi^+\pi^0 e^+e^-$
	&$2.93\,\text{fb}^{-1}\,\psi(3770)$
	&$1.4\times 10^{-5}$
	&\cite{Ablikim:2018gro}
	& --
	\\
$D^+\to K^+\pi^0 e^+e^-$
	&$2.93\,\text{fb}^{-1}\,\psi(3770)$
	&$1.5\times 10^{-5}$
	&\cite{Ablikim:2018gro}
	& --
	\\
$D^+\to K_S^0\pi^+ e^+e^-$
	&$2.93\,\text{fb}^{-1}\,\psi(3770)$
	&$2.6\times 10^{-5}$
	&\cite{Ablikim:2018gro}
	& --
	\\
$D^+\to K_S^0 K^+ e^+e^-$
	&$2.93\,\text{fb}^{-1}\,\psi(3770)$
	&$1.1\times 10^{-5}$
	&\cite{Ablikim:2018gro}
	& --
	\\
$D^0\to K^-K^+ e^+e^-$
	&$2.93\,\text{fb}^{-1}\,\psi(3770)$
	&$1.1\times 10^{-5}$
	&\cite{Ablikim:2018gro}
	&$3.15\times 10^{-4}$
	\\
$D^0\to\pi^+\pi^- e^+e^-$
	&$2.93\,\text{fb}^{-1}\,\psi(3770)$
	&$0.7\times 10^{-5}$
	&\cite{Ablikim:2018gro}
	&$3.73\times 10^{-4}$
	\\
$D^0\to K^-\pi^+ e^+e^-$
	&$2.93\,\text{fb}^{-1}\,\psi(3770)$
	&$4.1\times 10^{-5}$
	&\cite{Ablikim:2018gro}
	&$3.85\times 10^{-4}$
	\\
$D^0\to\pi^0 e^+e^-$
	&$2.93\,\text{fb}^{-1}\,\psi(3770)$
	&$0.4\times 10^{-5}$
	&\cite{Ablikim:2018gro}
	&$0.45\times 10^{-4}$
	\\
$D^0\to\eta e^+e^-$
	&$2.93\,\text{fb}^{-1}\,\psi(3770)$
	&$0.3\times 10^{-5}$
	&\cite{Ablikim:2018gro}
	&$1.1\times 10^{-4}$
	\\
$D^0\to\omega e^+e^-$
	&$2.93\,\text{fb}^{-1}\,\psi(3770)$
	&$0.6\times 10^{-5}$
	&\cite{Ablikim:2018gro}
	&$1.8\times 10^{-4}$
	\\
$D^0\to K_S^0 e^+e^-$
	&$2.93\,\text{fb}^{-1}\,\psi(3770)$
	&$1.2\times 10^{-5}$
	&\cite{Ablikim:2018gro}
	&$1.1\times 10^{-4}$
	\\
	\hline
$J/\psi\to D^0e^+e^-$
	&$1.31$B~$J/\psi$
	&$8.5\times 10^{-8}$
	&\cite{Ablikim:2017nid}
	&$1.1\times 10^{-5}$
	\\
	\hline
$\psip\to D^0e^+e^-$
	&$448$M~$\psip\;$
	&$1.4\times 10^{-7}$
	&\cite{Ablikim:2017nid}
	& --
	\\
$\psip\to \Lambda_c^+\bar{p}e^+e^-$~~~
	&$448$M~$\psip$~~~
	&$1.7\times 10^{-6}$
	&\cite{BESIII:2018ngn}
	& --
	\\
	\hline\hline
\end{tabular}
\label{tab:FCNC}
\normalsize
\end{table}

\subsection{Prospects for BESIII rare decay searches}
\label{subsect:rare}

The BESIII FCNC search results mentioned above are based on data collected in 2009-2012, which included
1.31B $\jpsi$ and 448M $\psip$ event samples and a 2.93~fb$^{-1}$ data sample that was accumulated at
$\ECM=3.773$~GeV, the peak energy of the $\psi(3770)\rt D\bar{D}$ resonance. BESIII has recently
increased the $\jpsi$ data sample to 10B events and will eventually increase the $\psip$ sample to
3B events, and the $\psi(3770)\rt D\bar{D}$ data to 20~fb$^{-1}$ (see Table 7.1 in
ref.~\cite{Ablikim:2019hff}). Since the results listed in Table~\ref{tab:FCNC} are mainly limited by
statistics, when the full data are available and analyzed, the sensitivity levels of FCNC searches
should improve, in most cases, by factors of $\sim 7$, and decay branching fractions will be probed at
the $10^{-6}$-$10^{-8}$ levels. If no interesting signals are found, more stringent upper limits would be
established that should further constrain the parameter spaces of a number of new physics models.

In contrast to FCNC processes, charged-current weak decays of charmonium states are allowed, but are
expected to occur as very rare processes; the SM-predicted branching fractions are of the order
$10^{-10}$-$10^{-8}$~\cite{SanchisLozano:1993ki}, which means they would be difficult to detect at
BESIII, even with the full 10B~event $\jpsi$ data sample. However, some BSM calculations based on a
two-Higgs-doublet model predict that the branching ratios of charmonium weak decays could be enhanced to
be as large as $10^{-5}$~\cite{Datta:1998yq}. BESIII searched for several Cabibbo-favored weak decays, such
as the hadronic processes $J/\psi\to D_s^-\rho^+$ and $J/\psi\to \bar{D}^0\bar{K}^{*0}$~\cite{Ablikim:2014dsn},
and the semi-leptonic process $J/\psi\to D_s^{(*)-}e^+\nu_e$~\cite{Ablikim:2014fpb}, and established 90\%~CL
branching fraction upper limits in the $\sim 10^{-5}$-$10^{-6}$ range. Searches for some Cabibbo-suppressed
weak decays of the $J/\psi$ are currently underway at BESIII, with expected branching fraction sensitivity
levels of about $10^{-7}$.

\section{Testing SM predictions for lepton couplings \& CKM matrix elements}

In the SM, the strength of charged-current weak interactions is governed by a single universal parameter, the
Fermi constant $G_{\rm F}$. The three charged leptons, ($e^-,\mu^-,\tau^-$) all couple to the $W$-boson with
this strength, a feature called lepton-flavor universality, LFU. Although the quarks appeared, at first, to
have different coupling strengths, this is because of a misalignment of
the  charge$=-1/3$ strong-interaction flavor eigenstates ($d,s,b$) and their weak-interactions counterparts
($d',s',b'$), as was first realized by Cabibbo in 1963~\cite{Cabibbo:1963yz}. He hypothesized that the weak
interaction flavor states were related to the strong-interaction states by an orthogonal rotation; the most
general rotation matrix for three quark generations was first written down by Kobayashi and Maskawa in
1973~\cite{Kobayashi:1973fv}. The universality of the quark-$W$ couplings is reflected by the unitarity of
the Cabibbo-Kobayashi-Maskawa (CKM) matrix. The equality of the weak interaction-coupling strengths for the
quarks and leptons is a feature that is specfic to the SM and is violated by many beyond-the-SM theories,
such as those that include fourth generation quarks, additional weak vector bosons, or multiple Higgs
particles.

\subsection{\boldmath Search for violations of charged lepton flavor universality  {(LFU)}}
\label{Sec:LFU}

The equality of the electron and muon couplings, $g_e$ and $g_{\mu}$, has been established at the
${\mathcal O}(0.2\%)$ level, i.e. $(g_e/g_{\mu} -1) = 0.002\pm 0.002$, by a comparison between the
$K^+\rt e^+\nu_e$ and $K^+\rt\mu^+\nu_{\mu}$ partial decay widths measured by the NA62
experiment~\cite{Lazzeroni:2012cx} together with PDG values for the $K^+$ lifetime and the electron \&
muon masses~\cite{Tanabashi:2018oca}.  The best test of the equality of the $\tau$-lepton coupling
and muon couplings,  $(g_{\tau}/g_{\mu} -1) = 0.0008\pm 0.0021$,  has similar precision and is from a
BESIII measurement of the tau mass~\cite{Ablikim:2014uzh} together with with PDG values of the
tau-lepton's lifetime and leptonic decay branching fractions.

The possibility of LFU violation has attracted considerable recent attention because of measurements from
BaBar~\cite{Lees:2012xj}, Belle~\cite{Abdesselam:2019dgh} and LHCb~\cite{Aaij:2017deq} of the
relative decay rates for the semileptonic processes $\bar{B}\rt D^{(*)}\tau^-\nu$
and  $\bar{B}\rt D^{(*)}\ell^-\nu$~($\ell^- = \mu^-\ {\rm or}\ e^- )$ that seem to violate SM expectations.
Specifically, the HFLAV Group's recent averages of experimental measurements are~\cite{Amhis:2019ckw}:
\footnotesize
\begin{eqnarray}
{\mathcal R}_D =  \frac{{\mathcal B}(\bar{B}\rt D\tau^-\nu)}{{\mathcal B}(\bar{B}\rt D\ell^-\nu)}&=&
0.340\pm 0.027\pm 0.013~({\rm expt.})~~[{\bf SM:}~~0.299\pm 0.003] \nonumber \\
{\mathcal R}_{D^*} =  \frac{{\mathcal B}(\bar{B}\rt D^{*}\tau^-\nu)}{{\mathcal B}(\bar{B}\rt D^{*}\ell^-\nu)}&=&
0.295\pm 0.011\pm 0.008~({\rm expt.)}~~[{\bf SM:}~~0.258\pm 0.005]. \label{eqn:RDstr}
\end{eqnarray}
\normalsize
Here the discrepancies with LFU, if they are real and not just statistical fluctuations, are of order
10\%, and motivate more careful checks of LFU in semileptonic and purely leptonic charmed particle decays with
BESIII data.

\subsubsection{BESIII tests of {LFU}}
Charmed particle decay measurements at BESIII are summarized in detail elsewhere in this volume~\cite{Li:2020xyz}.   
Table~\ref{tbl:lfu} summarizes measurements that are relevant for LFU tests, where all the
measurements agree with SM expectations within $1\sim 2\sigma$. The quantities in the last column,
$ \sqrt{(\Gamma_{e(\tau)}/\Gamma_\mu)/{\rm SM}}-1$, which would be $(g_{e(\tau)}/g_{\mu} -1)$ if radiative
corrections and detailed considerations of the relevant form-factors were properly applied, are included
as indicators of the sensitivity levels. According to these values, the most
stringent BESIII sensitivity levels for LFU-violating effects are a factor of five better than those
of the $\bar{B}\rt D^{(*)}\tau^-\nu$ measurements  (eqn.~\ref{eqn:RDstr}) but an order of magnitude
poorer than the limits on $g_e/g_{\mu}$ from $K^+$ decay.

\begin{table}[htbp]
  \footnotesize
  \caption{\footnotesize \label{tbl:lfu}  BESIII measurements of charmed particle semileptonic and
    purely leptonic branching-fraction measurements, and comparisons of the $\Gamma_{e(\tau)}/\Gamma_{\mu}$
    to SM expectations for LFU.}
    \centering
    \begin{tabular}{lcccccc}
 \hline 
 \hline 
 mode~~~&$n_{\rm evts}$ & ${\mathcal B}\ (\times 10^{-3})$ & ref. & $\Gamma_{e(\tau)}/\Gamma_\mu $ & SM~pred.~
                                                         &$ \sqrt{\frac{\Gamma_{e(\tau)}/\Gamma_\mu}{\rm SM}}-1$ \\
 \hline 
 $D^0\rt K^-\mu^+\nu_{\mu}$~~&~47.1K~&~~~$34.1\pm 0.2\pm 0.4$~~&\cite{Ablikim:2018evp}&
 ~~~\multirow{2}*{$1.027\pm 0.014$}~~~&~~~\multirow{2}*{$1.026\pm 0.001$}~
                                                         &~\multirow{2}*{$ 0.0\pm 0.007$} \\
 $D^0\rt K^- e^+\nu_e $      & 70.7K   & $35.05\pm 0.14\pm 0.03$  &    \cite{Ablikim:2015ixa} \\
 \hline 
   $D^0\rt\pim\mu^+\nu_{\mu}$   &  2.3K   & $2.72\pm 0.08\pm 0.06$  & \cite{Ablikim:2018frk}  &
 \multirow{2}*{$1.085\pm 0.037$} & \multirow{2}*{$1.015\pm 0.002$}
                                                         &~\multirow{2}*{$0.034\pm 0.019$}  \\
  $D^0\rt\pim e^+\nu_e $      & 6.3K   & $2.93\pm 0.04\pm 0.03$  &    \cite{Ablikim:2015ixa} \\         
 \hline 
 $D^+\rt\bar{K}^0\mu^+\nu_{\mu}$   &  20.7K   & $87.2\pm 0.7\pm 1.8$  & \cite{Ablikim:2016sqt}  &
                     \multirow{2}*{$1.012\pm 0.033$} & \multirow{2}*{$\approx 1.03$} & \\
 $D^+\rt \bar{K}^0 e^+\nu_e $      & 26.0K   & $86.0\pm 0.6\pm 1.5$  &    \cite{Ablikim:2017lks} \\
 \hline 
  $D^+\rt\piz\mu^+\nu_{\mu}$   &  1.3K   & $3.50\pm 0.11\pm 0.10$  & \cite{Ablikim:2018frk}  &
 \multirow{2}*{$1.037\pm 0.045$} & \multirow{2}*{$1.015\pm 0.002$}
                                                        &~\multirow{2}*{$0.011\pm 0.023$}  \\
  $D^+\rt\piz e^+\nu_e $      & 3.4K  & $3.63\pm 0.08\pm 0.05$  &   \cite{Ablikim:2017lks}  \\         
 \hline 
  $D^+\rt\omega\mu^+\nu_{\mu}$   &  194   & $1.77\pm 0.18\pm 0.11$  & \cite{BESIII:2020dbj}  &
                     \multirow{2}*{$0.92\pm 0.14$} & \multirow{2}*{$0.93 - 0.97$}&  \\
  $D^+\rt\omega e^+\nu_e $      &  491  & $1.63\pm 0.11\pm 0.08$  &   \cite{Ablikim:2015gyp}  \\         
 \hline 
  $D^+\rt\eta\mu^+\nu_{\mu}$   &  234   & $1.04\pm 0.10\pm 0.05$  & \cite{BESIII:2020dtz}  &
                     \multirow{2}*{$1.03\pm 0.13$} & \multirow{2}*{$1.0 - 1.03$} & \\
  $D^+\rt\eta e^+\nu_e $      &  373  & $1.07\pm 0.08\pm 0.05$  &   \cite{Ablikim:2018lfp}  \\         
 \hline 
  $\Lc\rt\Lambda\mu^+\nu_{\mu}$   &  1.3K   & $34.9\pm 4.6\pm 2.7$  & \cite{Ablikim:2016vqd}  &
                     \multirow{2}*{$1.04 \pm 0.31$} & \multirow{2}*{$\approx 1.0$} &  \\
  $\Lc\rt\Lambda e^+\nu_e $      & 104  & $36.3\pm 3.8\pm 2.0$  &   \cite{Ablikim:2015prg}  \\         
 \hline 
 \hline 
 $D^+\rt\tau^+\nu_{\tau}$   &  137   & $1.20\pm 0.24\pm0.12$  & \cite{Ablikim:2019rpl}  &
                     \multirow{2}*{$3.21\pm 0.77$} & \multirow{2}*{$2.67$} &~\multirow{2}*{$-0.09\pm 0.15$} \\
  $D^+\rt\mu^+\nu_{\mu} $      & 400  & $0.37\pm 0.02\pm 0.01$  &   \cite{Ablikim:2013uvu}  \\       
  \hline 
 $\Ds\rt\tau^+\nu_{\tau}$   &  22.1K   & $51.7\pm 1.5\pm 01.6$  & \cite{Hajime:2020abc}  &
                     \multirow{2}*{$9.42\pm 0.40$} & \multirow{2}*{$9.75$} & \multirow{2}*{$-0.02\pm 0.02$} \\
  $\Ds\rt\mu^+\nu_{\mu} $      & 1.1K  & $5.49\pm 0.16\pm 0.15$  &   \cite{Ablikim:2018jun}  \\       
 \hline 
 \hline 

    \end{tabular}
\normalsize
\end{table}

\subsubsection{Future prospects for {LFU} tests at BESIII}
The most stringent BESIII tests for LFU-violating effects in charmed-particle decays are derived
from measurements of $D\rt \bar{K}\ell^+\nu$  and $\pi\ell^+\nu$ semileptonic decays, where the
current $(g_e/g_{\mu} - 1)$ sensitivities are at the $1\sim 2\%$ level.  These results are based
on the analysis of the 2.93~fb$^{-1}$ data sample accumulated at the $\psi(3770)\rt D\bar{D}$ resonance.
When the analysis of the full 20~fb$^{-1}$ data set is complete,
the sensitivity levels of the LFU tests, which are now mostly statistically limited, will improve by
factors of $\sim 2.5$, and be in the sub-1\% range. In this case, if the current $1.8\sigma$
discrepancy that BESIII sees in $D^0\rt K^-\ell^-\nu$ is real and the central value reported in
Table~\ref{tbl:lfu} persists, its significance will increase to more than $4\sigma$. The other BESIII
measurement with interesting potential
is the ratio of the $\Ds\rt\tau^+\nu$ and $\Ds\rt\mu^+\nu$ purely leptonic decay rates that is based
on analyses of a 3.19~fb$^{-1}$ data sample collected at $\ECM=4178$~MeV, where
$\sigma(e^+e^-\rt \Dsstr \bar{D}_s^-)$ has a local maximum of $\sim$1~nb.  In this case, the BESIII
long-range plan includes  an additional 3~fb$^{-1}$ data sample at 4178~MeV, which would
provide a $\sqrt{2}$ improvement in $(g_{\tau}/g_{\mu} - 1)$ sensitivity.

\subsection{Unitarity of the CKM Matrix and the Cabibbo Angle Anomaly}
\label{Sec:CKM}

\begin{figure}[hbp]
\centering
  \includegraphics[height=0.18\textwidth,width=0.9\textwidth]{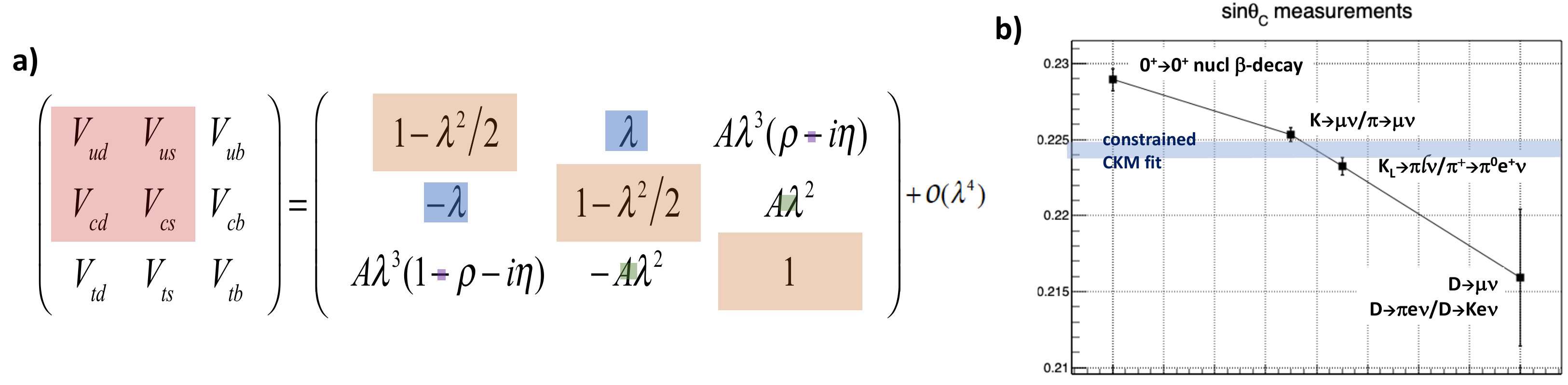}
  \caption{\footnotesize
    {\bf a)} The CKM matrix and its Wolfenstein parameterization. The shaded rectangles in
    the latter have areas $\propto |V_{ij}|$. {\bf b)} Values of $\sinthetac$ derived from
    different measurements.  The value based on nuclear $\beta$-decay is from~\cite{Seng:2018yzq},
    the one from $K_{\mu 2}$ ($K_{\ell 3}$) decays is from~\cite{Aoki:2019cca} (\cite{Bazavov:2018kjg}),
    and the one from $D$ decays is the average of BESIII
    ${\mathcal B}(D^+\rt\mu^+\nu)$~\cite{Ablikim:2013uvu} and
    ${\mathcal B}(D^0\rt \pim e^+\nu)/{\mathcal B}(D^0\rt K^- e^+\nu)$~\cite{Ablikim:2015ixa}
    measurements.  The shaded blue band is the PDG 2018 $\sinthetac$ value based on a
    unitarity-constrained  fit to all CKM elements~\cite{Tanabashi:2018oca}.
     }
    \label{fig:CKM-matrix}
\end{figure}

The CKM matrix (see Fig.~\ref{fig:CKM-matrix}a) is the DNA of flavor physics; its elements characterize
all of the SM weak charged current interactions of quarks. It defines a rotation in three-dimensions of
flavor-space and, in the SM where there are three quark generations, it must be exactly unitary; any
deviation from this would be a clear signal for new physics.

The unitarity condition for the top row of the CKM matrix is: $|V_{ud}|^2+|V_{us}|^2+|V_{ub}|^2 = 1$.
Experimentally, a high precision value of $|V_{ud}|$ comes from an analysis of eight superallowed
$0^+\rt 0^+$ nuclear $\beta$-decays~\cite{Hardy:2016vhg} corrected for electroweak effects. The
latest result is $|V_{ud}|=0.97370(4)$~\cite{Seng:2018yzq}.  A precise value of the
ratio $|V_{us}|/|V_{ud}|=0.2313(5)$ ratio is determined from a KLOE measurement of
${\mathcal B}(K^+\rt\mu^+\nu)$~\cite{Ambrosino:2005fw}, the PDG 2018 world average for
${\mathcal B}(\pip\rt\mu^+\nu)$~\cite{Tanabashi:2018oca} and a FLAG average of LQCD evaluations
of the pseudoscalar form-factor ratio $f_{K^+}/f_{\pip}$~\cite{Aoki:2019cca}. The value of $|V_{ub}|^2$,
determined from $B$-meson decays, is $\sim {\mathcal O}(10^{-5})$ and is a negligible contributor to
the unitarity condition~\cite{Tanabashi:2018oca}. The combination of these results~\cite{Seng:2018yzq},
\begin{equation}
  |V_{ud}|^2 +|V_{us}|^2 +|V_{ub}|^2 = 0.9983(5),
  \label{eqn:ckm}
\end{equation}
indicates a nominal $\sim 3.5\sigma$ deviation from unitarity that, if taken at face value, is strong
evidence for a SM violation.

Since deviations from CKM unitarity would be a clear sign of new physics, the eqn.~\ref{eqn:ckm}
result inspired further investigations. These included: independent determinations of $|V_{ud}|$ based
on the neutron lifetime~\cite{Czarnecki:2019mwq,Seng:2020wjq} that returned consistent results,
albeit with a slightly larger error; an independent evaluation of $|V_{us}|/|V_{ud}|$ using
${\mathcal B}(K_L\rt\pi\ell\nu)$ and ${\mathcal B}(\pip\rt\piz e^+\nu)$~\cite{Bazavov:2018kjg} that
found an even larger deviation from unitarity, but with a corresondingly larger error; and
reexaminations of the nuclear physics corrections used in the nuclear $\beta$-decay analyses for
$|V_{ud}|$~\cite{Seng:2018qru,Gorchtein:2018fxl} that did not change the central value, but indicated
that the previous error that was assigned to these effects may have been somewhat underestimated.
The current state of affairs is that the best current analyses of the existing data find an
${\mathcal O}(0.1\%)$ deviation from unitarity for the top row of the CKM matrix with a significance
level that is somewhere in the $2\sigma\sim 5\sigma$ range.

The strong generational hierarchy of the CKM quark-flavor mixing matrix is illustrated in
Fig.~\ref{fig:CKM-matrix}a, where the Wolfenstein parameterization~\cite{Wolfenstein:1983yz} is shown
with shaded rectangles with areas that are proportional to~$|V_{i,j}|$. Transitions between different
generations (i.e., further off-diagonal elements) are successively suppressed by additional factors of
$\lambda=\sinthetac \simeq 0.225$, where $\thetac$ is the Cabibbo angle. A striking feature of the
Wolfenstein formulation, and a characteristic of the SM, is that, to ${\mathcal O}(\lambda^6)\sim 10^{-4}$,
the four entries in the upper-left corner of the matrix, i.e., all transitions involving ($u,d$) \&
($c,s$) quarks, are well characterized by the single parameter, $\sinthetac$. The authors of
ref.~\cite{Grossman:2019bzp} argue that comparing the $\sinthetac$ values derived from different
$q_i\lrt q_j$  ($i=u,c;~j=d,s$) subprocesses is a more sensitive test for new physics than tests of
the CKM matrix unitarity, and provide, in support of this claim, an example of a toy model that has a
heavy gauge boson with different $d$- and $s$-quark couplings that demonstrates this. In
Fig.~\ref{fig:CKM-matrix}b, values of $\sinthetac$ derived from the nuclear
$\beta$-decay ($u\lrt d$) and $K_{\ell 2}~\&~K_{\ell 3}$ decays ($u\lrt s$) transitions discussed in the
previous paragraph are shown. The apparent discrepancy from a single, universal value is referred to
as the {\it Cabibbo angle anomaly}.

Studies of $c \rt d$ transitions provide  independent $\sinthetac$ determinations. In the
SM, $|V_{cd}|=|V_{us}|=\sinthetac$; a deviation between the $\sinthetac$ value inferred from $c\rt d$
decays with the one evaluated from  $K_{\ell 2}~\&~K_{\ell 3}$ decays would be another clear indication
of new physics. To date, this relation has not been strenuously tested. The PDG 2018 world-average
value, $|V_{us}|=0.2243\pm 0.0005$, differs from that for $|V_{cd}|=0.218\pm 0.004$ by $1.5\sigma$,
where the uncertainty of the latter is nearly an order of magnitude poorer~\cite{Tanabashi:2018oca}.
The best determinations of $|V_{cd}|$ to date are from statistically limited BESIII measurements
of ${\mathcal B}(D^+\rt\mu^+\nu)$~\cite{Ablikim:2013uvu} and the ratio
${\mathcal B}(D^0\rt \pim e^+\nu)/{\mathcal B}(D^0\rt K^- e^+\nu)$~\cite{Ablikim:2015ixa}, both of
which are based on analyses of BESIII's 2.97~fb$^{-1}$ sample of $\psi(3770)\rt D\bar{D}$ events
that are discussed elsewere in this volume~\cite{Li:2020xyz}. The average value of the two
$|V_{cd}|$ measurements is plotted in Fig.~\ref{fig:CKM-matrix}b. 

With the full 20~fb$^{-1}$ $\psi(3770)$  data sample, the BESIII precision on $|V_{cd}|$ should be
improved by at least a factor of 2.5; if the result is the same as the current
central value, the significance of the discrepancy would increase to about the $4\sigma$ level.

\section{\boldmath Searches for non-SM sources of $CP$ violation}
\label{Sec:CPV}

Searches for new sources of $CPV$ have been elevated to a new level of interest by the
recent LHCb discovery of a $CP$ violating asymmetry in the charmed quark sector; a
$5\sigma$ difference between the branching fractions for $D^0\rt K^+K^-$ or $\pi^+\pi^-$
and $\bar{D}^0$ to the same final states, with a magnitude of order~$10^{-3}$~\cite{Aaij:2019kcg};
the measured $CP$ violating asymmetry is at the high end of theoretical estimates for its SM
value, which range from $10^{-3}$~\cite{Golden:1989qx,Buccella:1994nf,Bianco:2003vb,Grossman:2006jg}
to $10^{-4}$~\cite{Khodjamirian:2017zdu}. Although the LHCb result is intriguing in that it may be a
sign of the long-sought-for non-SM mechanism for $CPV$, uncertainties in the SM calculations for this
asymmetry make it impossible to either establish or rule out this possibility~\cite{Saur:2020rgd}.

Violations of $CP$ have never been observed in weak decays of strange hyperons; the current limit
on $CPV$ asymmetry in $\Lambda$ hyperon decay is of order $10^{-2}$~\cite{Barnes:1996si}, which is
two orders-of-magnitude above the highest conceivable SM effects~\cite{Donoghue:1986hh}. A non-zero
measurement of a $CPV$ asymmetry at the level of $\sim 10^{-3}$ would be an unambiguous
signature for new physics.

\subsection{\boldmath Search for $CP$ violation in $\Lambda\rt p\pim$ decay}

\begin{figure}[hbp]
\centering
  \includegraphics[height=0.25\textwidth,width=1.0\textwidth]{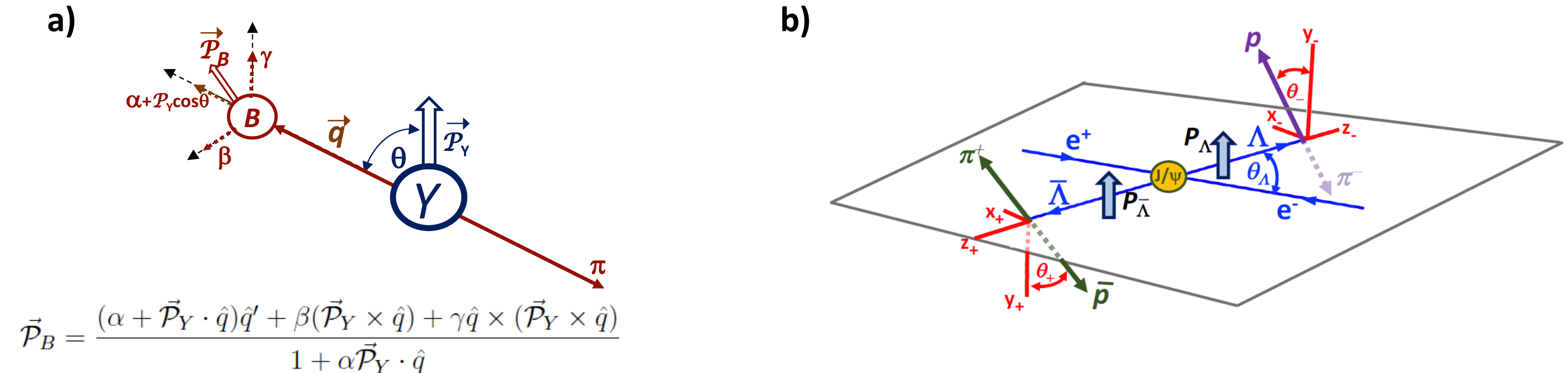}
  \caption{\footnotesize {\bf a)} Polarized $Y\rt B\pi$ decay illustrating 
    the $\alpha$, $\beta$, $\gamma$ dependence of the daughter $B$ polarization,
    where $\vec{q}$ is a vector along the $B$ momentum in the $Y$ rest frame. 
    {\bf b)} The $J/\psi\rt\Lambda\bar{\Lambda}$ reaction. Parity conservation in $J/\psi$
    decay guarantees that the ($\cos\theta$-dependent) $\Lambda$ and $\bar{\Lambda}$
    polarizations are equal and perpendicular to the production plane.
    }
    \label{fig:Lambda-decay}
\end{figure}

Parity violation in the weak interactions was discovered in 1957~\cite{Lee:1956qn,Wu:1957my}.
Immediately thereafter there was considerable interest is studying parity violations in strange
hyperon decays that were predicted by Lee and Yang~\cite{Lee:1957qs}.  For the $Y\rt B\pi$
weak decay process, where $Y$ is one of the spin $=1/2$ strange hyperons and $B$ is an octet
baryon, parity violation allows for both $S$- and $P$-wave transitions, and the final
states are characterized by the Lee-Yang parameters:

\footnotesize
\begin{equation}
  \alpha  = \frac{2{\rm Re}(S^*P)}{|S|^2+|P|^2};~~~\beta =\frac{2{\rm Im}(S^*P)}{|S|^2+|P|^2};
  ~~~\gamma =\frac{|S|^2-|P|^2}{|S|^2+|P|^2},
  \label{fig:abg}
\end{equation}
\normalsize
where $\alpha^2 +\beta^2 +\gamma^2 =1$. If the initial-state $Y$ has a non-zero
polarization~$\vec{\mathcal P}_Y$, the $B$ flight direction in the $Y$ rest frame relative to
the polarization direction, $\theta$, is distributed as
$dN/d\cos\theta \propto 1+\alpha|\vec{\mathcal P}_{Y}|\cos\theta$ and, if $\alpha$ is
also non-zero, has an explicit parity-violating up-down asymmetry.  The polarization of
the daughter baryon, $\vec{\mathcal P}_B$, depends on ${\mathcal P}_Y$, $\theta$, and the
$\alpha,\ \beta,\ \gamma$ parameters
as illustrated in Fig.~\ref{fig:Lambda-decay}a. If $CP$ is conserved, the decay parameters for
$Y$ and $\bar{Y}$ are equal in magnitude but opposite in sign. (The parameters for
$\bar{Y}$ are denoted by $\bar{\alpha}$ \& $\bar{\beta}$.) Violations of $CP$ symmetry would result
in non-zero values for the parameters $A_{CP}$ and $B_{CP}$, defined as
\footnotesize
\begin{equation}
  A_{CP}\equiv\frac{\alpha + \bar{\alpha}}{\alpha - \bar{\alpha}}~~~{\rm and}~~~
  B_{CP}\equiv\frac{\beta  + \bar{\beta}}{\beta - \bar{\beta}}.
  \label{fig:Acp}
\end{equation}
\normalsize
  
Measuring $\aLam$ for $\Lambda\rt p\pim$ decay is not straight forward.
Measurements of the up-down parity-violating asymmetry in $\Lambda\rt p \pim$ determine the product
$\aLam{\mathcal P}_{\Lambda}$, where ${\mathcal P}_{\Lambda}$ is generally unknown. To extract
$\aLam$, the polarization of the final-state proton must be measured. This was done in a series
of pre-1975 experiments by scattering the final-state proton on carbon, with a world-average result
of $\aLam=0.642\pm 0.013$~\cite{Bricman:1978ig}; this was the PDG value for 43 years, from 1976
until 2019.
  
BESIII  measured  $\aLam$ and $\aLamb$ with fully reconstructed
$\EE\rt\jpsi\rt (\Lambda\rt p\pim)(\bar{\Lambda}\rt\bar{p}\pip)$ events. For this reaction, the
joint angular distribution can be expressed as~\cite{Faldt:2017kgy}
\footnotesize
\begin{equation}
  d\Gamma\propto (1+\alpha_{\psi}\cos^2\theta_\Lambda)
  [1+{\mathcal P}_{\Lambda}(\cos\theta_{\Lambda})(\aLam\cos\theta_- +\aLamb\cos\theta_+)]
  +\aLam\aLamb[{\mathcal F}_1(\xi)+({1-\alpha_{\psi}^2})^{\frac{1}{2}}\cos\Delta\Phi{\mathcal F}_2(\xi)],
\label{eqn:dsdxi}
\end{equation}
\normalsize
where: $\theta_{\Lambda}$ is the $\Lambda$ production angle relative to the $e^+$-beam direction (the
$\cos\theta_{\Lambda}$ distribution is $1+\alpha_{\psi}\cos^2\theta_{\Lambda}$); $\Delta\Phi$ is the complex
phase difference between the $A_{+,+}$ and $A_{+,-}$ helicity amplitudes; and $\xi$ denotes
$(\theta_{\Lambda},\theta_-,\phi_-\theta_+,\phi_+)$, where $\theta_- ,\phi_-$ ( $\theta_+ ,\phi_+$)
are the $\Lambda$ ($\bar{\Lambda}$) decay angles (see Fig.~\ref{fig:Lambda-decay}b). The
$\cos\theta_{\Lambda}$-dependent $\Lambda$ (and $\bar{\Lambda}$) polarization is given by
\footnotesize
\begin{equation}
  {\mathcal P}_{\Lambda}(\cos\theta_{\Lambda})=\frac{
    ({1-\alpha_{\psi}^2})^{\frac{1}{2}}\cos\theta_{\Lambda}\sin\theta_{\Lambda}\sin\Delta\Phi}
               {1+\alpha_{\psi}\cos^2\theta_{\Lambda}}.
\end{equation}
\normalsize
The $\Lambda$ polarization is zero if the $A_{+,+}$ and $A_{+,-}$ helicity amplitudes are relatively real
(i.e., $\Delta\Phi=0$), in which case it is apparent from eqn.~\ref{eqn:dsdxi} that only the product
$\aLam \aLamb$ can be measured and individual determinations of $\aLam$ and $\aLamb$ cannot
be extracted from the data. (Expressions for ${\mathcal F}_1(\xi)$ and ${\mathcal F}_2(\xi)$ are provided
in ref.~\cite{Faldt:2017kgy}.)

\begin{figure}[hbp]
\centering
  \includegraphics[height=0.2\textwidth,width=0.75\textwidth]{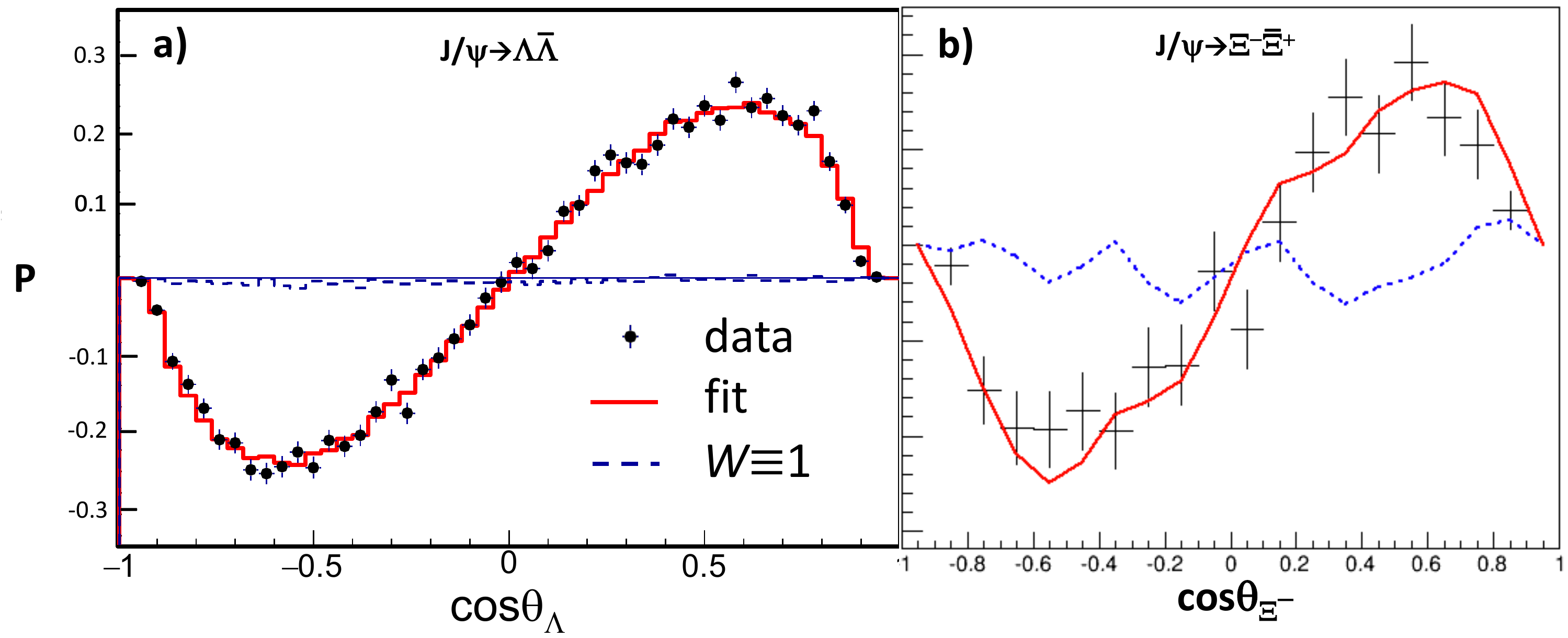}
  \caption{\footnotesize Polarization {\it vs} $\cos\theta_{\Lambda (\Xi^-)}$ for
    {\bf a)} $J/\psi\rt\Lambda\bar{\Lambda}$~\cite{Ablikim:2018zay} and {\bf b)}
    $J/\psi\rt\Xi^-\bar{\Xi}^+$~[BESIII preliminary~\cite{Adlarson:2019abc}]
    events. The red curves are fits to the data; the blue curves are expectations for
    zero polarization. 
    }
    \label{fig:pol}
\end{figure}

When BESIII was being planned, it was generally thought that ${\mathcal P}_{\Lambda}\approx 0$ and that
$J/\psi\rt\Lambda\bar{\Lambda}$ events would not be useful for $CP$ tests. It was somewhat of a surprise
when BESIII subsequently discovered that, in fact, the polarization of $\Lambda$ and $\bar{\Lambda}$
hyperons produced in $\jpsi$ decays is substantial~\cite{Ablikim:2018zay}, as shown in
Fig.~\ref{fig:pol}a.   With a sample of 420K fully reconstructed
$J/\psi\rt (\Lambda\rt p\pim)(\bar{\Lambda}\rt\bar{p}\pip)$ events in a 1.31B~$J/\psi$ event sample,
BESIII measured  $A_{CP}^{\Lambda}=-0.006\pm 0.012\ \pm 0.007$. This null result improved on the precision
of the best previous measurement, $A_{CP}^{\Lambda}=+0.013 \pm 0.022$~\cite{Barnes:1996si}, that was based
on 96K $p\bar{p}\rt\LLB$ events, by a factor of two.  As a by-product of this measurement, BESIII made
the world's most precise measurement of  $\aLam =0.750\pm 0.010$, a result that is more than five standard
deviations higher than the previous PDG average value. It is likely that all previous measurements
were  biased by a common systematic problem, probably related to the spin analyzing properties of carbon;
the PDG 2019 value for $\aLam$ is solely based on the BESIII value~\cite{Tanabashi:2018oca}.

\subsection{\boldmath Prospects for BESIII $CP$ violation studies}

The BESIII values for $A_{CP}^{\Lambda}$ and $\aLam$ mentioned in the previous paragraph were
realized by an analysis of 1.3\ B $J/\psi$ decays, which is a small subset of BESIII's
total 10B $J/\psi$ event sample. The analysis of the full data set is currently
underway that, when done, will provide a factor-of-three improvement in sensitivity.

BESIII is currently applying a similar analysis to
$J/\psi\to (\Xi^-\rt\Lambda\pim)(\bar{\Xi}^+\rt\bar{\Lambda}\pip)$ hyperon pairs, where preliminary
results demonstrate that there is substantial transverse $\Xi$ polarization (see
Fig.~\ref{fig:pol}b). In $\Xi^-\bar{\Xi}^+$ events, the $\alpha_{\Xi}$ decay parameter influences
both the up-down decay asymmetry in the primary $\Xi\rt\Lambda\pi$ process, and the polarization
of the daughter $\Lambda$ hyperons (see Fig.~\ref{fig:pol}a) that can be determined from
the decay asymmetry in the secondary $\Lambda\rt p\pim$ decay. For a given sample of $J/\psi$
decays, the number of fully reconstructed $\Xi^-\bar{\Xi}^+$ events in which $\Lambda\rt p\pim$ and
$\bar{\Lambda}\rt\bar{p}\pip$ are only about one quarter of the number of reconstructed
$\jpsi\rt\Lambda\bar{\Lambda}$ events because of the smaller $J/\psi\rt\Xi^-\bar{\Xi}^+$ branching
fraction and a lower detection efficiency. Nevertheless, this lower event number is compensated by
the added information from the daughter $\Lambda$ decays. As a result, the sensitivity per event for
the $\Xi^-$ decay parameters is higher than that for $\Lambda$ parameters with $\jpsi\rt\LLB$ events, and
simulations show comparable precisions for $\alpha_{\Xi^-}$ and $\alpha_{\Lambda}$~\cite{Adlarson:2019jtw}.
In contrast to $\Lambda\rt p\pi$, where measuring the daughter proton's polarization is impractical, in
$\Xi\rt\Lambda\pi$ decays the daughter $\Lambda$ polarization is measured and $B_{CP}^{\Xi^-}$ can be
determined; $B_{CP}^{\Xi^-}$ is potentially more sensitive to new physics than
$A_{CP}^{\Xi^-}$~\cite{Gonzalez:1994zc}.

In addition to the $\Lambda$ hyperons produced by $J/\psi\rt\Lambda\bar{\Lambda}$, those produced as
daughters in $J/\psi\to (\Xi^-\rt\Lambda\pim)(\bar{\Xi}^+\rt\bar{\Lambda}\pip)$ events are also useful
for $A_{CP}^{\Lambda}$ measurements. The rms polarization of $\Lambda$ hyperons produced via $\jpsi\rt\LLB$
(see Fig.~\ref{fig:pol}a) is $\langle {\mathcal P}_{\jpsi,\Lambda}\rangle_{\rm rms}\approx 0.13$. In contrast,
the rms polarization for $\Lambda$ hyperons produced as a daughter particle in $\Xi^-\rt\Lambda\pim$
decay is $\langle {\mathcal P}_{\Xi^-,\Lambda}\rangle_{\rm rms}\approx|\alpha_{\Xi^-}|=0.39\pm0.01$ (see
Fig.~\ref{fig:Lambda-decay}a). Thus,
$\langle {\mathcal P}_{\Xi^-,\Lambda}\rangle_{\rm rms}\approx 3\langle {\mathcal P}_{\jpsi,\Lambda}\rangle_{\rm rms}$
and, since the $A_{CP}^{\Lambda}$ sensitivity is proportional to $\sqrt{n_{\rm evts}}$ but linear in
$\langle {\mathcal P}_{\Lambda}\rangle_{\rm rms}$, a $\Lambda$ from $\Xi^-\rt\Lambda\pim$ decay has nine
times the equivalent statistical power of a $\Lambda$ from $J/\psi\rt\Lambda\bar{\Lambda}$. Detailed
estimates of BESIII's ultimate statistical error for $A_{CP}$ with the existing 10B $\jpsi$ event sample,
including $\Lambda$ hyperons from $\Xi\rt\Lambda\pi$ decays, are reported in ref.~\cite{Adlarson:2019jtw}
and summarized here in Table~\ref{tbl:sens}. The projected ultimate $A_{CP}^{\Lambda}$ sensitivity
is ${\mathcal O}(2\times 10^{-3})$, which is an order of magnitude improvement on the pre-BESIII
result~\cite{Barnes:1996si}. 

\begin{table}[htbp]
    \footnotesize
  \caption{\footnotesize  \label{tbl:sens} The expected numbers of fully reconstructed events and the
    extrapolated $1\sigma$ statistical errors on $\langle\alpha\rangle = (\alpha - \bar{\alpha})/2$ and
    $A_{CP}$ from a complete analysis of $\jpsi\rt\LLB$,~$\XmXmB$~and $\XzXzB$ events in BESIII's 10B
    $\jpsi$ event data sample (from ref.~\cite{Adlarson:2019jtw}). Here the full reconstruction of
    the $\Lambda\rt p\pim$ and $\bar{\Lambda}\rt\bar{p}\pip$ decay channels are required. 
  }
 \centering
 \begin{tabular}{lcccccccccccc}
 \hline 
 \hline 
 reaction~~~~~~~& ${\mathcal B}\ (\times 10^{-4})$ &  $n_{\rm evts}$  &~~~&
 ~~~$\delta\AL$~~&~~$\delta\ALCP$~~&~~~~&~~$\delta\AXim$~~&~~$\delta\AXimCP$~~&~~~~&~~$\delta\AXiz$~~&~~$\delta\AXizCP$~~\\
 \hline 
 $\jpsi\rt\LLB$   &   18.9    &    3,200K   &~~~~&
 0.0010     &    0.0049          &   &                 &                    &   &                &                  \\
 $\jpsi\rt\XmXmB$ &   9.7     &   810K      &~~~&  
 0.0018     &    0.0034          &   &      0.0016     &     0.0039         &   &                &                  \\
 $\jpsi\rt\XzXzB$ &   11.6    &   670K      &~~~&  
 0.0019     &    0.0041          &   &                 &                    &   &      0.0017    &     0.0049        \\
  \hline 
  combined        &    &    & &
  0.0013     &    0.0023          &   &                 &                    &   &                &                  \\
  \hline 

 \end{tabular}
 \normalsize
\end{table}

\section{Standard model forbidden processes}

Cross sections for $e^+e^-\rt {\it hadrons}$ in the BESIII accessible $\ECM$ regions 
are ${\mathcal O}(10~{\rm nb})$ and the experiment typically records ${\mathcal O}(10^5)$
events/day. However, at the $J/\psi$ resonance peak, the cross section
is $\approx 3.6\mu$b, and in a typical day of operation BESIII collects ${\mathcal O}(10^8)$~events.
The cross section at the $\psip$ peak is $\approx 2\mu$b and the event rate
is ${\mathcal O}(5\times 10^7)$~events/day.  Thus, at the $\jpsi$ and $\psip$ peaks, BESIII has a high rate
of events in a very clean experimental environment that are well suited for high sensitivity searches for
a number of SM-model forbidden processes. About one third of the $\psip$ events decay via $\psip\rt\ppjpsi$,
where the triggering on, and detection of only the $\pp$ pair provides an unbiased ``beam'' of tagged
$\jpsi$ mesons that can be used to search for decays to final states that would otherwise be undetectable. 

\subsection{\boldmath Search for the Landau-Yang theorem forbidden $\jpsi\rt\gamma\gamma$ decay}

The Landau-Yang theorem states that a massive spin~1 meson cannot decay to two
photons~\cite{Landau:1948kw,Yang:1950rg}. As a consequence, the $\jpsi\rt\gamma\gamma$ decay mode is
strictly forbidden. An unambiguous signal for $\jpsi\rt\gamma\gamma$ would signal a breakdown of the
spin-symmetry theorem of QFT, the underlying framework of the SM and its many proposed
new physics extensions.  (For a discussion of how QFT might be modified to accommodate a
Landau-Yang theorem violation see ref.~\cite{Gninenko:2011ws}.)

  \begin{figure}[hbp]
\centering
  \includegraphics[height=0.2\textwidth,width=0.75\textwidth]{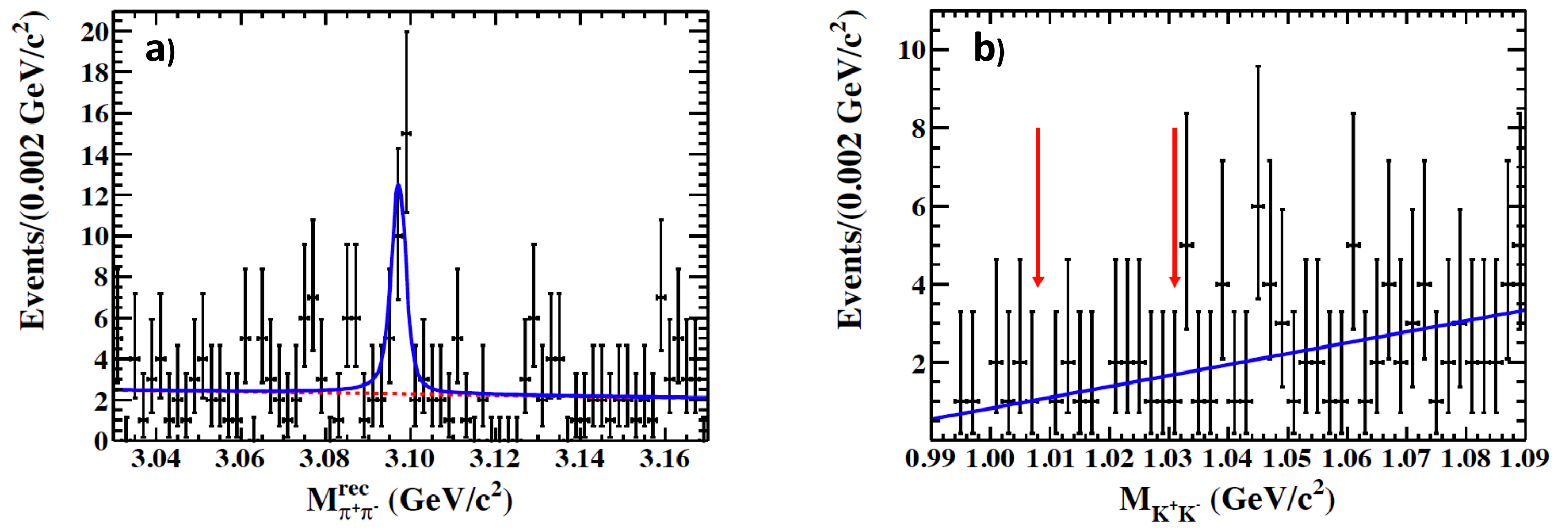}
  \caption{\footnotesize 
    {\bf a)} The $\pp$ recoil mass spectrum for selected  $\psip\rt\pp\gamma\gamma$ events.
    The peak at the $\pp$ recoil mass$\approx m_{\jpsi}=3.097$~GeV is entirely attributable to
    backgrounds from $\jpsi\rt\gamma\piz$ and $\gamma\eta$. 
     {\bf b)} The $K^+K^-$ invariant mass distribution for $\psip\rt \gamma\pp K^+K^-$ events with
    $M(\gamma K^+K^-)=m_{\jpsi}\pm 15$~MeV. A $\jpsi\rt\gamma\phi$ decay would show up as a narrow
    peak with $M(K^+K^-)\approx m_{\phi}=1.02$~GeV.  Both plots are from ref.~\cite{Ablikim:2014viz}.
    }
    \label{fig:jpsigg}
\end{figure}

The PDG~2018 upper limit, ${\mathcal B}(\jpsi\rt\gamma\gamma)<1.6\times 10^{-7}$~\cite{Tanabashi:2018oca},
is entirely based on a BESIII measurement that uses tagged $\jpsi$ mesons that recoil from the $\pp$
system in $\psip\rt\pp\jpsi$ decays~\cite{Ablikim:2014viz}, and is a factor of 20~times more sensitive
than previous measurements. In a data sample containing 106M $\psip$ decays, events with two oppositely
charged tracks and two $\gamma$-rays that satisfy a four-constraint energy-momentum kinematic fit
to the $\pp\gamma\gamma$ hypothesis were selected. Figure~\ref{fig:jpsigg}a shows the mass recoiling
against the $\pp$ tracks where there is a  $29\pm 7$~event peak at the $\jpsi$ mass that is consistent
with being entirely due to the expected background from roughly equal numbers of $\jpsi\rt\gamma\piz$
and $\gamma\eta$ events in which the $\piz$ and $\eta$ decay to a pair of $\gamma$-rays with a large
energy asymmetry and the low energy $\gamma$ is undetected either because its energy is below the
detection threshold or it outside of the fiducial acceptance region of the detector
($|\cos\theta_{\gamma}|>0.92$).

\subsubsection{ Search for the $C$-parity violating $\jpsi\rt\gamma\phi$ decay}

A similar BESIII analysis searched for $\jpsi\rt\gamma\phi$~\cite{Ablikim:2014viz}. Although
this process does not violate the Landau-Yang theorem, it violates charge-parity ($C$) conservation.
The weak interactions are known to violate $C$-parity, but the expected branching fractions for
weak-interaction-mediated $\jpsi$ decays are below the level of $10^{-9}$~\cite{Wang:2016dkd}. 
If $\jpsi\rt\gamma\phi$ were seen with a branching fraction that is higher than this, it would be
imply a $C$-parity violation in the electromagnetic interaction and be an indicator of new physics. This
measurement is based on a search for $\jpsi$ decays to $\gamma\phi$; $\phi\rt K^+K^-$, with tagged $\jpsi$
mesons from $\psip\rt\pp\jpsi$ decays. In this case kinematically constrained $\gamma\pp K^+K^-$ events,
where the $K^+$ and $K^-$ are positively identified as such by the BESIII pid systems and the $\pp$ recoil
mass is within $\pm 15$~MeV of $m_{\jpsi}$.  Figure~\ref{fig:jpsigg}b shows the $K^+K^-$
invariant mass where there is no sign of a $\phi\rt K^+K^-$ peak at $M_{K^+K^-}\approx m_{\phi}=1020$~MeV.
A 90\% CL upper limit on the size of the $\phi$ signal is $<6.9$~events, which translates into a
branching fraction upper limit of ${\mathcal B}(\jpsi\rt\gamma\phi)<1.4\times 10^{-6}$. This is the
first experimental limit for this decay.

\subsection{\boldmath Search for lepton flavor violation in $\jpsi\rt e\mu$ decays}

The discovery of neutrino oscillations~\cite{Fukuda:1998mi} provided clear evidence for violations of
lepton flavor conservation (LFV) in the neutrino sector.  However, the SM translation of the neutrino
results to the charged-lepton sector predicts LFV effects that are proportional to powers of the neutrino
masses with branching fractions that are immeasurably small ($< 10^{-51}$).  Thus, any observation of LFV at
levels  much higher than this would be clear evidence for new physics, such as grand unified (GUT) theories
or the presence of extra dimensions. Although most attention is given to LFV searches in muon decay, tau
decay and $\mu\rt e$ conversion experiments, in some theories LFV quarkonium decays, including
$V\rt \ell^-_i\ell^+_j$ decays, where $i\neq j$, are promising reactions~\cite{Bordes:2000gd}. BESIII
searched for the LFV decay $\jpsi\rt e^-\mu^+$.

  \begin{figure}[hbp]
\centering
  \includegraphics[height=0.20\textwidth,width=0.9\textwidth]{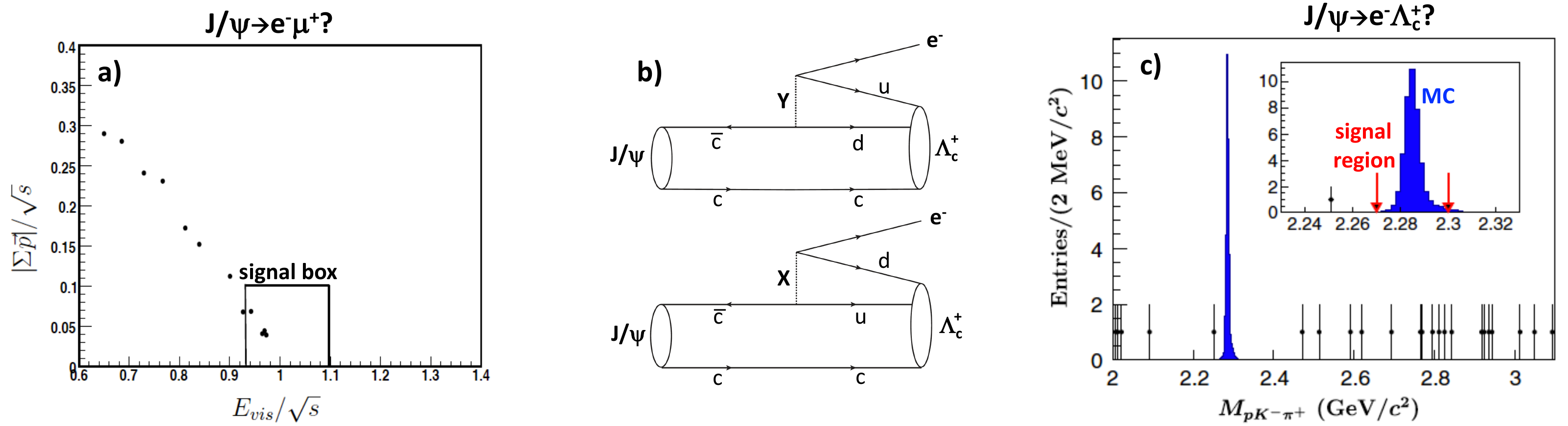}
  \caption{\footnotesize 
    {\bf a)} $|\sum \vec{p}|/\sqrt{s}$ {\it vs.} $E_{\rm vis}/\sqrt{s}$ for selected  $J/\psi\rt e^-\mu^+$
    candidate events in BESIII ~\cite{Ablikim:2013qtm}. {\bf b)} Diagrams for leptoquark-mediated
    $\jpsi\rt e^-\Lc$ decay as per the model of ref.~\cite{Pati:1974yy}. {\bf c)} The $pK^-\pip$ invariant
    mass distribution for selected, kinematically constrained  $J/\psi\rt e^-p K^-\pip$ events
    (from BESIII~\cite{Ablikim:2018uiq}).  The expected shape of a $\jpsi\rt\Lc e^-$; $\Lc\rt p K^-\pip$
    signal is shown as the blue histogram.
    \label{fig:LFV}
    }
\end{figure}

  The best previous limit was a 2003 BESII result,
  ${\mathcal B}(\jpsi\rt e^-\mu^-)<1.1\times 10^{-6}$ \cite{Bai:2003sv},
that was based on an analysis of a sample of $58$M $\jpsi$ events. This was improved by a 2013 BESIII
result that used a sample of 230M $\jpsi$ events. In this analysis, the variables $|\sum \vec{p}|/\sqrt{s}$
and $E_{\rm vis}/\sqrt{s}$ are examined for events with two back-to-back and oppositely charged tracks, with
one track positively identified as an electron and the other as  muon. Events with detected $\gamma$-rays or
additional tracks are rejected, and selected events are required to satisfy a four-constraint energy-momentum
kinematic fit.  The main background is expected to be from $\jpsi\rt\mu^+\mu^-$ events in which one of the
muons passes the electron identification requirements.  Figure~\ref{fig:LFV}a  shows a scatterplot of
$|\sum \vec{p}|/\sqrt{s}$ {\it vs.} $E_{\rm vis}/\sqrt{s}$ for selected events, where the four events in
the signal box are consistent with the $4.75\pm 1.09$ background events that are expected. (This
background level corresponds to a muon to electron misidentification probability of $\sim 10^{-7}$.)
The 90\%~CL upper limit of ${\mathcal B}(\jpsi\rt e^-\mu^+)<1.6\times 10^{-7}$ that is
established~\cite{Ablikim:2013qtm} is a factor of seven more stringent than the previous result.

\subsection{\boldmath Search for lepton/baryon number violations in $\jpsi\rt \Lambda_c^+ e^-$}

In addition to $CPV$, another requirement that Sakharov listed for the production of the matter-antimatter
symmetry of the universe is the existence of a mechanism for baryon/lepton number
violation~\cite{Sakharov:1967dj}. Processes that violate baryon (\textsc{B}) and lepton number (\textsc{L})
but conserve their difference (\textsc{B-L}) occur in GUT theories~\cite{Pati:1974yy}. Experiments that
search for \textsc{B}-violating decays of the proton have reported lifetime upper limits with spectacular
sensitivities: e.g., $\tau(p\rt e^+\piz)> 1.6\times 10^{34}$~years~\cite{Miura:2016krn}. In contrast, limits
for \textsc{B}-violating decays in the heavy quark sector are sparse and not remotely as sensitive. These
include a 90\% CL upper limit ${\mathcal B}(D^0\rt pe^-)< 1.0\times 10^{-5}$ from CLEO~\cite{Rubin:2009aa}
and BaBar branching fraction limits for $B^0\rt\Lc\ell^-$ and $B^-\rt \Lambda(\bar{\Lambda})\ell^-$ (here
$\ell= e,\mu$) that range from a ${\rm few}\times 10^{-6}$ for the $\Lc$ modes to a ${\rm few}\times 10^{-8}$
for the $\Lambda (\bar{\Lambda})$ modes~\cite{BABAR:2011ac}.

The only result on \textsc{B}-violating quarkonium decays is a BESIII upper limit on $\jpsi\rt\Lc e^-$ that
is based on an analysis of a sample of 1.31B $\jpsi$ decays. Quark line diagrams for this process in the
context of the Pati-Salam model~\cite{Pati:1974yy} are shown in Fig.~\ref{fig:LFV}b, where \textsc{X} and
\textsc{Y} are virtual leptoquarks that mediate the decay. BESIII search searched for exclusive
$\jpsi\rt \Lc e^-$ decay events where the $\Lc$ decays to $ pK^-\pip$ (${\mathcal B}=6.3\%$). The $pK^-\pip$
invariant mass distribution for candidate events, shown as data points in Fig.~\ref{fig:LFV}c, has no events
in the mass interval that is $\pm 4$~times the resolution and centered on the $\Lc$ mass. The absence of any
event candidates translates into a 90\% CL frequentist upper
limit of ${\mathcal B}(\jpsi\rt\Lc e^-)<6.9\times 10^{-8}$~\cite{Ablikim:2018uiq}.

\section{Searches for New, Beyond the Standard Model Particles}

In spite of the success of the SM, particle physics still faces a number of mysteries and challenges,
including the origin of elementary particle masses and the nature of dark matter (DM). The Higgs
mechanism~\cite{Higgs:1964pj} is a theoretically attractive way to explain the mass of elementary
particles. However, the SM relation for the Higgs mass is a potentially divergent infinite sum of
quadratically increasing terms that somehow add up to the finite value $m_{\rm Higgs}=125$~GeV, a
SM feature that many theoretical physicists consider to be {\it unnatural}~\cite{Susskind:1978ms}. The
existence of DM is inferred from a number of astrophysical and cosmological observations~\cite{Jibrail:2019slb}.
One possibility is that DM may be comprised of electrically neutral, weakly interacting, stable particles
with a mass at the electroweak scale. However, none of the SM particles are good DM candidates and, from the
perspective of theory and phenomenology, this implies that the SM is deficient and the quest for a more
fundamental theory beyond the SM is strongly motivated. In some extensions of the SM, the naturalness and
DM problems can be solved at once. 

The naturalness problem can be solved by supersymmetry (SUSY)~\cite{Martin:1997ns}, where every SM particle
has an as yet undiscovered partner with the same quantum numbers and gauge interactions but differs in spin
by $\frac{1}{2}$. The most economical and intensively studied version of SUSY is the minimal supersymmetric
model (MSSM)~\cite{Martin:1997ns}, with superpartners that include:
\begin{center}
\begin{tabular}{lll}
spin-zero~~~       & \emph{sfermions}:~~~ & left handed $\tilde{f}_L$,  right handed $\tilde{f}_R$;\\
spin-$\frac{1}{2}$ & \emph{gauginos}:  & a bino $\tilde{B}$, three winos $\tilde{W}_i$, gluinos $\tilde{g}$;\\
spin-$\frac{1}{2}$ & \emph{higgsinos}: & two $\tilde{H}_i$.
\end{tabular}
\end{center}
The two higgsinos can mix with the bino and the three winos to produce two \emph{chargino} $\chi_{1,2}^\pm$
and four \emph{neutralino} $\chi_{1,2,3,4}^0$ physical states. A discrete symmetry called $R$-parity is
introduced to make the lightest SUSY particle, usually the $\chi_1^0$, stable, which makes it a nearly ideal
DM candidate that is often denoted as simply $\chi$. A further extension is the so-called next-to-minimal MSSM
(NMSSM)~\cite{Ellwanger:2009dp,Maniatis:2009re,Djouadi:2008uw}, in which a complex isosinglet field is added.
The NMSSM has a rich Higgs sector containing three $CP$-even, two $CP$-odd, and two charged Higgs bosons. The
mass of the lightest $CP$-odd scalar Higgs boson, the $A^0$, may  be less than twice the mass of charm quark,
in which case it would be accessible at BESIII. 

Although the lightest neutralino is an attractive DM candidate, the lack of any experimental evidence for it
in either LHC experiments or direct detection experiments suggests that DM might be more complex than the
neutralino of the SUSY models. Attempts to devise a unified explanation have led to a vast and diverse array
of dark-sector models. These models necessarily have several sectors: a {\em visible sector} that includes all
of the SM particles, a {\em dark sector} of particles that do not interact with the known strong, weak, or
electromagnetic forces, and a {\em portal sector} that consists of particles that couple the visible and dark
sectors. The latter may be vectors, axions, higgs-like scalars or neutrino-like
fermions~\cite{Essig:2013lka,Alexander:2016aln}, of which vectors are the most frequently studied. The simplest
scenario for the vector portal invokes a new force that is mediated by a $U(1)$ gauge boson~\cite{Holdom:1985ag}
that couples very weakly to charged particles via kinetic mixing with the SM photon $\gamma$, with a mixing
strength $\varepsilon$ that is in the range between $10^{-5}$ and $10^{-2}$~\cite{ArkaniHamed:2008qn}. This new
boson is variously called a dark photon, hidden photon or $U$~boson, and is denoted as $\gamma'$. The $\gamma'$
mass is expected to be low, on the order of MeV/$c^2$ to GeV/$c^2$~\cite{ArkaniHamed:2008qn} and, thus, it could
be produced at the BEPCII collider in a variety of processes, depending on its mass.

\subsection{\boldmath Search for $A^0$, $\gamma'$ and invisible decays of light mesons}

Both the light $CP$-odd NMSSM Higgs boson $A^0$ and dark photon $\gamma'$ have been searched for by BESIII. Since
it is Higgs-like, the $A^0$ couples to SM fermions with a strength proportional to the fermion mass. For an
$A^0$ with a mass below the $\tau$ pair production threshold, the decay $A^0\to\mu^+\mu^-$ is expected to be
dominant. The $A^0$ can also serve as a portal to the dark sector with the invisible-final-state decay 
process $A^0\to\chi\bar\chi$.  Similarly, as a portal between the SM and dark sectors, the $\gamma'$
can, in turn, either decay to $\chi\bar\chi$, or visibly to a pair of light leptons or
quarks, provided it is kinematically allowed.

\begin{figure}[hbp]
\centering
  \includegraphics[width=0.5\textwidth]{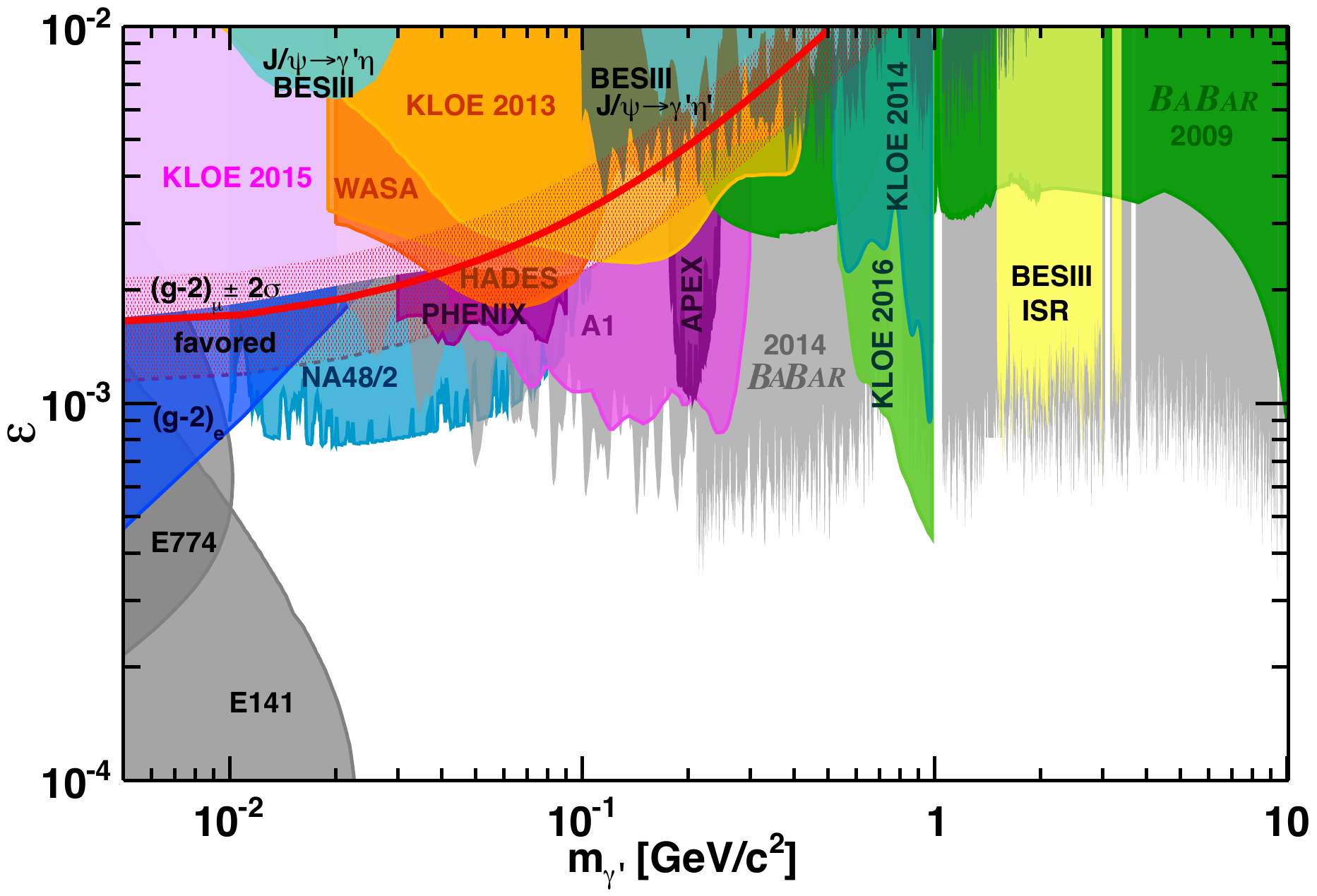}
  \caption{\footnotesize Exclusion limits at the 90\% confidence level for the mixing strength parameter
    $\varepsilon$ as a function of the dark photon mass $m_{\gamma'}$. Also shown are exclusion limits
    from other experiments. The $\varepsilon$ values that would explain the discrepancy between the measured
    and SM-calculated value of the anomalous magnetic moment of the muon~\cite{Pospelov:2008zw} are displayed
    as the bold solid red line along with its $2\sigma$ band.
    }
    \label{fig:DarkPhoton_EMDalitz_MixPar}
\end{figure}

BESIII results on searches for the $A^0$, $\gamma'$ and invisible decays of light meson states are summarized
in Table~\ref{tab:bsm}. The $A^0$ was searched for in $J/\psi\to\gamma A^0$~($ A^0\to\mu^+\mu^-$) and
$\psip\to\pi^+\pi^-\jpsi$~($\jpsi\to\gamma A^0$)~($A^0\to\mu^+\mu^-$) decay candidate events in BESIII's
$\jpsi$~\cite{Ablikim:2015voa} and $\psip$~\cite{Ablikim:2011es} data samples. The sensitivity obtained
with the $\jpsi$ data is five times better than that with the $\psip$ data. The combination of
BaBar~\cite{Lees:2012iw} and BESIII~\cite{Ablikim:2015voa} measurements constrain the $A^0$ to be mostly
singlet. BESIII published three results on dark photon ($\gamma'$) searches in $\jpsi$ and $\psi(3770)$
decays with resulting 90\% CL exclusion regions for $\varepsilon$ as a function of the dark
photon mass that are shown in Fig.~\ref{fig:DarkPhoton_EMDalitz_MixPar}. 
  BESIII dark photon searches in $\jpsi\to\eta\gamma'$~($\gamma'\to e^+e^-$) decays~\cite{Ablikim:2018eoy}
  and $\jpsi\to\eta'\gamma'$ ($\gamma'\to e^+e^-$) decays~\cite{Ablikim:2018bhf} were among the first
  searches that were based on these channels~\cite{Ablikim:2017aab}. BESIII results for dark photon
  searches in $e^+e^-\to\gamma_\text{ISR}\gamma'(\gamma'\to\ell^+\ell^-, \ell=e,\mu)$ initial state
  radiation events were based on two years of data taking and are competitive with BaBar
  results~\cite{Lees:2014xha} based on nine years of running.
Invisible decays of light mesons produced  $\jpsi$ decays were also searched for at BESIII. These include
the first measurements for the $\omega$ and $\phi$ vector mesons that are copiously produced via
$\jpsi\to\omega\eta$ and $\phi\eta$ decays~\cite{Ablikim:2018liz}. For
$\jpsi\to\phi\eta$~($\eta\to\emph{invisible}$)and $\jpsi\to\phi\eta'$~($\eta'\to\emph{invisible}$) decays,
the BESIII limits~\cite{Ablikim:2012gf} are factors of six and three improvements over previous results
from BESII~\cite{Ablikim:2006eg}. These results provide complementary information to studies of the nature
of DM and constrain parameters of the phenomenological models.

\normalsize

\begin{table}[htbp]
\footnotesize
  \caption{\footnotesize BESIII results on searches for light $CP$-odd Higgs boson $A^0$, dark photon
    $\gamma'$, and invisible decays of quarkonium and light mesons. The first column lists the decay modes and the
    third column lists the measured 90\%~CL branching fractions upper limits. For the visible dark photon
    decays, the corresponding $\gamma-\gamma'$ mixing strength $\varepsilon$ limits are shown in the
    fourth column.}
\label{tab:bsm}
\begin{tabular}{llllll}
\hline\hline
Mode  		&Data 
			&$\mathcal{B}^\text{UL}$ at 90\% CL
			&$\varepsilon(\times10^{-3})$~~~~		
			&ref. \\
\hline
$J/\psi\to\gamma A^0 (\to\mu^+\mu^-)$ 
			&$225$M~$\jpsi$
			&$(2.8-495.3)\times 10^{-8}$
			&
			&\cite{Ablikim:2015voa}\\
$\psi'\to\pi\pi\jpsi(\to \gamma A^0 (\to\mu^+\mu^-))$~~ 
			&$106$M~$\psip$ 
			&$(4-210)\times 10^{-7}$ 
			&
			&\cite{Ablikim:2011es}\\
\hline
$J/\psi\to \eta \gamma'(\to e^+e^-)$&\multirow{2}*{$1.310$B~$\jpsi$}
			&$(1.9-91.1)\times 10^{-8}$~~~~~~~
			&$10-1$
			&\cite{Ablikim:2018eoy}\\ 
$J/\psi\to \eta' \gamma'(\to e^+e^-)$
			&
			&$(1.8-20)\times 10^{-8}$
			&$3.4 - 26$
			&\cite{Ablikim:2018bhf}\\ 
$e^+e^-\to \gamma_\text{ISR}\gamma'(\to e^+e^-/\mu^+\mu^-)$
			&$2.93\,\text{fb}^{-1}\ \psi(3770)$~~~~~
			&
			&$0.1 - 1$
			&\cite{Ablikim:2017aab}\\ 
\hline
$\jpsi\to \eta\omega(\omega\to\emph{invisible})$
			& \multirow{2}*{$1.31$B~$\jpsi$}
			&$7.3\times 10^{-5}$
                        & & \multirow{2}*{\cite{Ablikim:2018liz}}\\
$\jpsi\to \eta\phi(\phi\to\emph{invisible})$
			&
			& $1.7\times 10^{-4}$
			& &\\
\hline
$\jpsi\to \phi\eta(\eta\to\emph{invisible})$
			& \multirow{2}*{$225$M~$J/\psi$}
			&$1.0\times 10^{-4}$
			&& \multirow{2}*{\cite{Ablikim:2012gf}}\\ 
$\jpsi\to \phi\eta'(\eta'\to\emph{invisible})$
			&
			&$5.3\times 10^{-4}$
			&& \\ 
\hline\hline
\end{tabular}
\normalsize
\end{table}

\section{Interactions with Other Experiments}

The standard model of particle physics is a seamless structure in which measurements in
one sector have profound impact on other, seemingly unrelated areas. Thus, for example,
BESIII measurements of strong-interaction phases in hadronic decays of charmed mesons provide
important input into determinations of the $CP$-violating angle $\gamma$ in $B$-meson decays
by BelleII and LHCb. Similarly, BESIII measurements of the annihilation cross section for
$\EE\rt hadrons$ at energies below 2~GeV provide critical input to the interpretation of
high energy tests of the SM at the Higgs (126~GeV) and top-quark(173 GeV) mass scales as
well as the measurements of $(g-2)_\mu$, the anomalous magnetic moment of the muon. The
relation between BESIII measurements of strong phases in the charmed sector to $CPV$
measurements in the beauty sector are discussed elsewhere in this volume~\cite{Li:2020xyz}. Here we
briefly review the impact of BESIIII cross section results on the interpretation of
$(g-2)_\mu$ measurements.

\subsection{\boldmath BESIII impact on the determination of $(g-2)_\mu$ }

\begin{figure}[htbp]
  \centering
  \includegraphics[height=0.22\textwidth,width=0.8\textwidth]{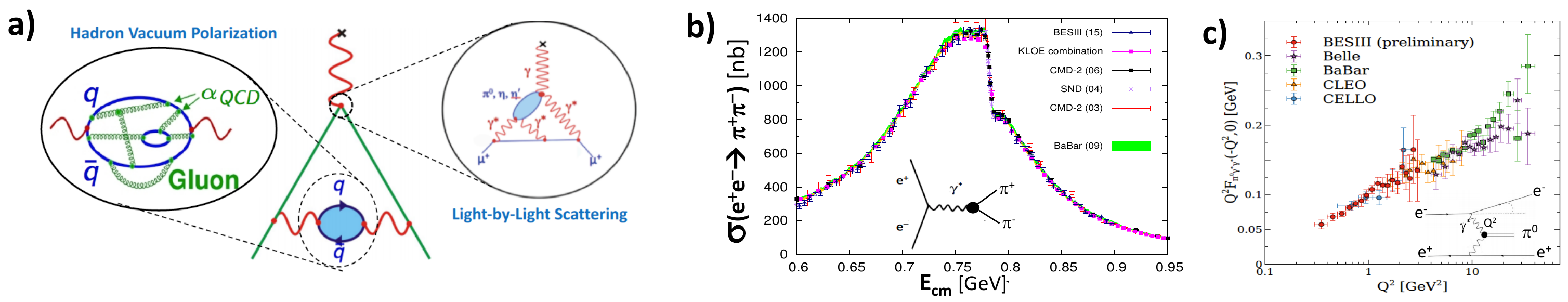}
  \caption{\footnotesize
    {\bf a)} Hadron vacuum polarization (HVP) and hadron light-by-light scattering (HLbL) contributions
    to the SM calculation of $(g-2)_\mu$.
    {\bf b)}   Measurements of $\sigma(e^+e^-\to \pp)$ from SND~\cite{Achasov:2006vp},
    CMD-2~\cite{Akhmetshin:2003zn,Akhmetshin:2006bx}, BaBar~\cite{Lees:2012cj}, KLOE~\cite{Anastasi:2017eio},
    and BESIII~\cite{Ablikim:2015orh}. The structure near $\ECM=0.78$~GeV is caused by interference between
    $\rho\to \pp$ and $\omega\to \pp$ (from ref.~\cite{Keshavarzi:2018mgv}).
    {\bf c)} Preliminary BESIII results for the $\pi^0$ form factor~\cite{Redmer:2018uew} together with
    results from CELLO~\cite{Behrend:1990sr}, CLEO~\cite{Gronberg:1997fj}, BaBar~\cite{Aubert:2009mc}
    and Belle\cite{Uehara:2012ag} (from ref.~\cite{Aoyama:2020ynm}).
    }    
  \label{fpi}
\end{figure}

The measured value of $(g-2)_\mu$ from BNL experiment E821~\cite{Bennett:2006fi} 
is $\sim$3.7 standard deviations higher than the SM prediction~\cite{Aoyama:2020ynm},
a discrepancy that has inspired elaborate follow-up experiments at
Fermilab~\cite{Grange:2015fou} and J-PARC~\cite{Otani:2015jra}. As illustrated in
Fig.~\ref{fpi}a, the SM predicted value for $(g-2)_{\mu}$ is very sensitive to the effects
of hadronic vacuum polarization (HVP)
of the virtual photon, which are about 100 times larger than the current
experimental uncertainty. The contributions from higher-order radiative corrections to the
$\mu$-$\gamma$~vertex, so-called hadron light-by-light (HLbL) scattering, is of the same order
as the current experimental error, but it has a 20\% theoretical uncertainty that will be
comparable to the expected error from the new round of experiments.

Vacuum polarization also has critical influence on precision tests of the
electroweak theory that rely on a precise knowledge of $\alpha(s)$, the
running QED coupling constant. Because of vacuum polarization,
$\alpha^{-1}(m^2_Z)=128.95\pm 0.01$~\cite{Davier:2017zfy}, about 6\%
below its long-distance value of $\alpha^{-1}(s=0)=137.04$. About half of this
change is due to HVP.

\subsubsection{Precision measurement of vacuum polarization of virtual photons}

Since HVP effects are non-perturbative, they cannot be directly computed from first
principle QCD. Recent computer-based Lattice QCD (LQCD) calculations have made significant 
progress but the uncertainties are still large~\cite{Miura:2019xtd,Davies:2019efs}. 
The most reliable determinations to date of HVP contributions
to $(g-2)_\mu$ and $\alpha(m^2_Z)$ use dispersion relations with input from
experimental measurements of cross sections for $\EE$ annihilation into
hadrons~\cite{Aoyama:2020ynm}.  
The data used for the most recent determinations
are mostly from the SND~\cite{Achasov:2006vp},
BaBar~\cite{Lees:2012cj}, BESIII~\cite{Ablikim:2015orh},
CMD-2~\cite{Akhmetshin:2003zn,Akhmetshin:2006bx}, and KLOE~\cite{Anastasi:2017eio} experiments.
BaBar and KLOE operations have been terminated, 
leaving SND, CMD-3~\cite{Akhmetshin:2016dtr}, and BESIII as the
only running facilities with the capability to provide the improvements in precision
that will be essential for the evaluation of $(g-2)_\mu$ with a precision that
will match those of the new experimental measurements.

With data taken at $\ECM=3.773$~GeV (primarily for studies
of $D$-meson decays) BESIII measured the cross sections for $\EE\to\pp$
at $\ECM$ between 0.6 and 0.9~GeV~\cite{Ablikim:2015orh}, which covers the $\rho\to\pp$
peak, the major contributor to the HVP dispersion relation integral.
These measurements used initial state radiation (ISR) events in which
one of the incoming beam particles radiates a $\gamma$-ray
with energy $E_{\rm ISR}=x\ECM /2$ before annihilating at a reduced CM
energy of $\ECM=\sqrt{1-x}\ECM$. The relative uncertainty of
the BESIII measurements is 0.9\%, which is similar to the precision of 
the BaBar~\cite{Lees:2012cj} and KLOE~\cite{Anastasi:2017eio} results. 
The BESIII measured values agree well with KLOE results
for energies below 0.8~GeV but are systematically higher
at higher energies; in contrast, BESIII results agree with BaBar
at higher energies but are lower at lower energies. 
Detailed comparisons are shown in Fig.~\ref{fpi}b.
Nevertheless, the contributions of $\EE\to \pp$ to the $(g-2)_{\mu}$ HVP calculation
from these experiments have overall agreement within two standard deviations, and the
observed $\sim$3.7 standard deviation difference between the calculated muon
magnetic moment value and the E821 experimental measurement persists.

\subsubsection{Experimental input for data-driven HLbL determinations}

The HLbL scattering contribution to the SM $(g-2)_\mu $ value, has a hadron loop (see Fig.~\ref{fpi}a)
that is non-perturbative and in a more complex environment than the HVP loop. As a result,
its determination is not straightforward and has a rather volatile history (see
ref.~\cite{Melnikov:2016wdt}). In this case, the loop integral is dominated by single mesons
($\pi^0$, $\eta$, $\eta'$) but, since they couple to virtual photons, their time-like form factors at
low $Q^2$ values are involved. Until now, only high $Q^2$ measurements of these form factors have been
reported and models were used to extrapolate these to the low $Q^2$ regions of interest. Recently,
however, BESIII reported preliminary $\pi^0$ form-factor results for $Q^2$ values in the range
$0.3$-$1.5$~GeV$^2$~\cite{Redmer:2018uew} (see Fig.~\ref{fpi}c). These are the first experimental results
that include momentum transfers below $Q^2=0.5$~GeV$^2$, the relevant region for HLbL calculations. These,
and measurements of the $\eta$ and $\eta'$ form factors that are currently underway, will reduce the
model dependence and, thus, the theoretical errors of the HLbL contribution to $(g-2)_\mu$. 

\subsection{\boldmath Prospects for $(g-2)_\mu$-related measurements at BESIII}

Currently, the precision of the  $(g-2)_\mu$ measurement (54~ppm~\cite{Bennett:2006fi})
is comparable to that of the SM calculation (37~ppm~\cite{Aoyama:2020ynm}).
However, since a four-fold improvement in the experimental precision is imminent,
improvements in the theoretical precision are needed. These will require improved experimental input
for the data-driven evauations of the HVP and HLbL terms and/or
improved LQCD calculations. BESIII is improving the  $\sigma(\EE\rt hadrons)$ measurements
used for the HVP term and providing light-meson form-factors
for the HLbL determination. Moreover, precision BESIII measurements of various
decay constants and form-factors provide calibration points that are used to validate LQCD techniques.


\section{Summary and Perspectives}
\label{Sec:Summary}
In the search for new, beyond the standard model physics, there is no compelling
theoretical guidance for where it might first show up. It may first appear at the energy
frontier that is explored at the LHC, or at the intensity frontier that is pursued at
lower energies. (Interestingly, the current most prominent candidate for BSM physics
is the $\sim 3.7\sigma$ discrepancy in $(g-2)_\mu$, which is about as far removed from the
energy frontier as one can get.) A key aspect of any experiment is {\em reach}, i.e. the
range of unexplored SM-parameter space that is explored. In this quest, BESIII
is accumulating huge numbers of $\jpsi$ and $\psip$ events that
support high sensitivity searches for low-mass non-SM particles, SM-forbidden decay
processes, and non-SM $CP$ violations in hyperon decays.  In addition, high statistics
samples of $D$ and $D_s$ mesons produced just above threshold in very clean experimental
environments provide the means to search for new physics in the ($u,d$)-($c,s$) quark
sector with world's best precision. BESIII is
continuing the  BES program's long history of steadily improving the precision of
$\EE\rt~hadrons$ annihilation cross section measurements and light meson form factor
determinations that are used to evaluate HVP and HLbL corrections that are needed for the
interpretation of SM tests being done by other experiments.

\section*{Acknowledgments}
\noindent
This work is supported in part by 
National Key Basic Research Program of China under Contract No. 2015CB856700; 
Joint Large-Scale Scientific Facility Funds of the NSFC and CAS under Contract No. U1532257;
the CAS President’s International Fellowship Initiative; and the Korean Institute for Basic
Science under project code IBS-R016-D1.

\bibliography{NP_references}
\end{document}